\documentclass{PoS}

\usepackage{color}
\let\normalcolor\relax
\usepackage{rotating}
\usepackage{textcomp}

\definecolor{orange}{rgb}{1.0,.6,0}
\definecolor{green}{rgb}{0,0.7,0} 

\newcommand{\tbr}{{\color{red}$\blacksquare$}} 
\newcommand{\tbg}{{\color{green}$\bigstar$}}
\newcommand{\tbo}{{\color{orange}\Large$\bullet$}}
\newcommand{\gA}{{\color{green} \tt\bf{A}}}
\newcommand{\oP}{{\color{orange} \tt\bf{P}}} 
\newcommand{\rC}{{\color{red} \tt\bf{C}}}

\newcommand{\bea}{\begin{eqnarray}}
\newcommand{\eea}{\end{eqnarray}}
\newcommand{\beq}{\begin{equation}}
\newcommand{\eeq}{\end{equation}}

\newcommand{\mev}{{\rm MeV}}

\newcommand{\GeV}{\,\mathrm{GeV}}
\newcommand{\msbar}{{\overline{{\rm MS}}}}

\title{Kaon physics from lattice QCD}

\ShortTitle{Kaon physics from lattice QCD}

\author{\speaker{Vittorio Lubicz} 
        \\
        Universit\`a Roma Tre and INFN sezione di Roma Tre\\
        E-mail: \email{lubicz@fis.uniroma3.it}}

\abstract{
I review lattice calculations and results for hadronic parameters relevant for
kaon physics, in particular the vector form factor $f_+(0)$ of semileptonic kaon
decays, the ratio $f_K/f_\pi$ of leptonic decay constants and the kaon bag
parameter $B_K$. For each lattice calculation a colour code rating is assigned,
by following a procedure which is being proposed by the Flavianet Lattice
Averaging Group (FLAG), and the following final averages are obtained: $f_+(0) =
0.962(3)(4)$, $f_K/f_\pi = 1.196(1)(10)$ and $\hat B_K = 0.731(7)(35)$. In the
last part of the talk, the present status of lattice studies of non-leptonic $K
\to \pi\pi$ decays is also briefly summarized.
}

\FullConference{The XXVII International Symposium on Lattice Field Theory -
LAT2009\\
		 July 26-31 2009\\
		 Peking University, Beijing, China}

\begin{document}

\section{Introduction}
Kaon physics has always played a crucial role for our understanding of
fundamental interactions in the flavour sector. Together with B-physics, for
which however precise experimental information has become available only in the
last decade, the study of kaon physics has allowed fundamental tests of the
Standard Model and provides, at present, precise constraints on its possible new
physics extensions.

While there are few selected processes in kaon physics, like the rare $K \to \pi
\nu \bar \nu$ decays, which can be studied with almost negligible theoretical
uncertainties, in most of the cases the extraction of the physical results also
relies on our capability of controlling the non-perturbative effects of the
strong interactions and, therefore, on the accuracy of lattice QCD calculations.
This is the case, for example, of the determination of the CKM matrix element
$V_{us}$ from the study of semileptonic and leptonic kaon decays, or the
theoretical prediction of the $\varepsilon_K$ parameter which controls the
amount of indirect CP violation in $K^0-\bar{K}^0$ mixing. The hadronic parameters
entering these processes, namely the vector form factor $f_+(0)$ of semileptonic
kaon decays, the leptonic decay constant $f_K$ and the kaon bag parameter $B_K$,
are the quantities to be computed in lattice QCD calculations. 

In the last few years, the accuracy of the lattice predictions for kaon physics
observables is significantly increased. Extensive unquenched lattice simulations
have been performed by various lattice collaborations, using different lattice
methods (i.e. different actions, renormalization techniques, etc.). A list of
these simulations, their main details and a compilation of kaon observables
which have been studied with them, is presented in table~\ref{tab:kaonphys}.
\begin{table}
\begin{center}
\renewcommand{\arraystretch}{1.8}
\begin{tabular}{cccccc}
Collaboration & Quark action & $N_f$ & $a$ [fm] & \parbox{1.1cm}{\centering 
$m_\pi^{\rm min}$ \\[-1.0mm]{[}MeV]} & \parbox{1.8cm}{\centering Kaon \\[-1.0mm]
observables} \\[2.0mm] \hline
\parbox{3.7cm}{\centering MILC \\[-1.0mm] {[}+ FNAL, HPQCD, \ldots ]} 
& \parbox{1.6cm}{\centering Improved \\[-1.0mm] Staggered} & 2+1 & $\ge 0.045$ &
230 & $f_K$, $B_K$  \\
PACS-CS & Clover (NP) & 2+1 & 0.09 & 156 & $f_K$  \\
RBC/UKQCD & DWF & 2+1 & $\ge 0.08$ & 295 & \parbox{2.3cm}{\centering $f_+(0)$,
$f_K$, $B_K$, \\[-1.0mm] $K \to \pi\pi$} \\
BMW & \parbox{1.6cm}{\centering Clover \\[-1.0mm] Smeared} & 2+1 & $\ge 0.07$ &
190 & $f_K$  \\
JLQCD/TWQCD & Overlap & \parbox{0.9cm}{\centering 2 \\[-1.0mm] {[}2+1]} & 0.12 &
290 & $B_K$ \\
ETMC & \parbox{1.6cm}{\centering Twisted \\[-1.0mm] Mass} &
\parbox{1.3cm}{\centering 2 \\[-1.0mm] {[}2+1+1]} & $\ge 0.07$ & 260
& $f_+(0)$, $f_K$, $B_K$ \\
QCDSF & Clover (NP) & 2 & $\ge 0.07$ & 340 & $f_K$ \\[2.0mm] \hline
\end{tabular}
\renewcommand{\arraystretch}{1.0}
\end{center}
\caption{Details of the unquenched lattice simulations which have been used for
the studies of kaon physics. The relevant kaon observables computed in these
calculations are listed in the last column.}
\label{tab:kaonphys}
\end{table}
These simulations typically involve ensembles with different lattice spacings,
which allow the continuum extrapolation to be eventually performed. They also
include a number of relatively light simulated quark masses, with the lightest
pion masses now typically smaller than 300 MeV. In this mass region, a
controlled chiral extrapolation to the physical light quark masses can be
performed, using in most of the cases chiral perturbation theory (ChPT) as a
useful guideline for the extrapolation.

The relative abundance of unquenched lattice calculations, in which most (if not
all) the systematic uncertainties are kept well under control and which use
different approaches, characterized by different systematics, has allowed to
reach in the determination of the corresponding hadronic parameters a
significantly improved accuracy. It should be also noted that a table similar to
table~\ref{tab:kaonphys} for lattice studies of B-physics would present, today,
a much shorter list.

In this talk I will mainly concentrate on reviewing the recent lattice results
for the three hadronic parameters which are of particular interest for kaon
physics, namely the vector form factor $f_+(0)$ of semileptonic kaon decays, the
ratio $f_K/f_\pi$ of leptonic decay constants and the kaon bag parameter $B_K$.
I will also mostly rely, for this task, on the detailed work which is being
performed by the FLAG group, that will be introduced in the next section. For
each lattice calculation, a colour code rating in the FLAG style will be
assigned, and I will also present my best averages for the hadronic parameters.
I will conclude this talk by briefly summarizing the status of lattice studies
of non-leptonic $K \to\pi\pi$ decays.

\section{The FLAG working group: rating and averaging criteria for lattice
results}

The improved control of systematic uncertainties achieved in the last few years
by lattice QCD calculations, particularly for observables in the kaon sector,
simplifies the task of deriving the corresponding lattice averages to be
used in phenomenological analysis. This kind of task is one of those which is
currently being addressed by the Flavianet Lattice Averaging Group
(FLAG)~\cite{flag}, a working group of the Flavianet European network
constituted in November 2007.

The aim of FLAG is to provide for each considered quantity, to the network's
working groups and to the wider community, the following information: i) a
collection of current lattice results and references; ii) a summary of the
essential aspects of each calculation; iii) averages of lattice results.

The quantities which are being considered in the first FLAG report~\cite{flag}
are the light quark masses ($m_u$, $m_d$, $m_s$), the SU(2) and SU(3) low energy
constants, the kaon semileptonic form factor $f_+(0)$, the ratio of leptonic
decay constants $f_K/f_\pi$ and the bag parameter $B_K$. It is clear from this
list that a significant overlap exists between the FLAG work and the task I have
been given at this conference. I will then take advantage of this overlap, and I
will use for this talk several FLAG results. I am indebted and grateful for that
to my colleagues in the FLAG group.

One of the FLAG proposals that I'm going to follow in this review concerns the
way of summarizing the essential aspects of each lattice calculation. This is
done by using an easy-to-read ``colour code'' classification. Specifically, a
number of sources of systematic errors are identified and a colour with respect
to each of these is assigned to each calculation. The prescription is as
follows:
\begin{itemize}
\vspace{-0.25cm}
\item[\tbg] when the systematic error has been estimated in a
satisfactory manner and convincingly shown to be under control;
\vspace{-0.25cm}
\item[\tbo] when a reasonable attempt at estimating the systematic error has
been made, although this could be improved;
\vspace{-0.25cm}
\item[\tbr] when no or a clearly unsatisfactory attempt at estimating the
systematic error has been made.
\end{itemize} 
\vspace{-0.25cm}

It should be clear that the precise criteria used in determining the colour
coding are unavoidably time-dependent. The sources of systematic error and
definitions which are currently adopted by FLAG, and which I'm also going to
follow in this talk, are:
\begin{itemize}
\vspace{-0.2cm}
\item \underline{Chiral extrapolation}:\\
\tbg \hspace{0.2cm}  $M_{\pi,\mathrm{min}}< 250$ MeV  \\
\tbo \hspace{0.2cm}  250 MeV $\le M_{\pi,{\mathrm{min}}} \le$ 400 MeV \\
\tbr \hspace{0.2cm}  $M_{\pi,\mathrm{min}}> 400$ MeV \\
It is assumed that the chiral extrapolation is done with at least a three-point
analysis. In case of nondegeneracies among the different pion states $M_\pi$
stands for an average pion mass.

\vspace{-0.2cm}
\item \underline{Continuum extrapolation}:\\
\tbg \hspace{0.2cm}  3 or more lattice spacings, at least 2 points below
0.1 fm\\ 
\tbo \hspace{0.2cm}  2 or more lattice spacings, at least 1 point below 0.1
fm \\ 
\tbr \hspace{0.2cm}  otherwise\\
It is assumed that the action is $O(a)$-improved. The colour coding criteria for
non-improved actions change as follows: one lattice spacing more needed. 

\vspace{-0.2cm}
\item \underline{Finite-volume effects}:\\
\tbg \hspace{0.2cm}  $(M_\pi L)_\mathrm{min} > 4$ or at least 3 volumes \\
\tbo \hspace{0.2cm}  $(M_\pi L)_\mathrm{min} > 3$ and at least 2 volumes \\
\tbr \hspace{0.2cm}  otherwise\\
It is assumed that $L_\mathrm{min}\ge $ 2 fm, otherwise a red dot will be
assigned. In case of nondegeneracies among the different pion states $M_\pi$
stands for an average pion mass.

\vspace{-0.2cm}
\item \underline{Renormalization} (where applicable):\\
\tbg \hspace{0.2cm}  non-perturbative\\
\tbo \hspace{0.2cm}  2-loop perturbation theory (with a converging series) \\
\tbr \hspace{0.2cm}  otherwise

\vspace{-0.2cm}
\item \underline{Running} (where applicable): \\
\tbg \hspace{0.2cm}  non-perturbative\\
\tbo \hspace{0.2cm}  otherwise\\
\tbr \hspace{0.2cm}  ---
\end{itemize}
\vspace{-0.2cm}

Of course any colour coding has to be treated with caution and it goes without
saying that these criteria are subjective and evolving. Moreover, the extent to
which each source of systematic uncertainty affects the lattice calculation is
observable dependent. FLAG believes, however, that this attempt to introduce
quality measures for lattice results, in spite of being necessarily schematic,
will prove to be a useful guide.

The other main purpose of FLAG is to provide averages of lattice results. The
average should only include, as far as possible, only ``good quality'' lattice
calculations. This is implemented, in practice, by relying on the colour coding:
unless special reasons are given for making an exception, the averages are
restricted to data for which the colour code does not contain any red dot. In
deriving the averages quoted in this talk, I will follow the same criterium.

There are two other criteria adopted by FLAG for computing the averages which,
however, I'm not going to apply for the purposes of the present review. One is
related to the publication status, for which a coloured symbol is also
introduced:
\begin{itemize}
\vspace{-0.2cm}
\item \underline{Publication status}:\\
\gA \hspace{0.2cm}  published or plain update of published results  \\
\oP \hspace{0.2cm}  preprint \\
\rC \hspace{0.2cm}  conference contribution
\end{itemize}
\vspace{-0.2cm}
The FLAG policy is to consider in the averages only calculation which have been
published, i.e. which have been endorsed by a referee. While I find this policy
perfectly justified for the FLAG purposes, I also believe that the same
criterium is not suitable for the reviewer at the lattice conference. The latter
is asked to concentrate the attention mainly on the new (and typically
unpublished) results presented at the conference. It is the task of the
reviewer, rather than of an external referee in this case, to judge the quality
and the reliability of the presented results.

The other criterium adopted by FLAG is related to the number of flavours of
dynamical quarks introduced in the simulation. The policy that is being followed
by FLAG consists in presenting separate averages for the $N_f=2$ and $N_f=2+1$
calculations. This issue has been quite debated within the working group, and I
personally do not consider the choice currently pursued by FLAG an optimal one
for a review of lattice results. There are mainly three reasons for that, in my
opinion, that I would like to mention here, since the issue is also relevant for
the present review.

i) I believe that it would be useful to present, to the wider community of
particle physics, the ``best lattice result'' in terms of just a
single number, rather than two. When separate averages for the $N_f=2$ and
$N_f=2+1$ results are quoted, the natural interpretation of the latter as the
best result may not always correspond to the actual situation. The error due to
the quenching of the strange quark is rather small in most of the cases, whereas
other sources of systematic uncertainty could be better under control in the
$N_f=2$ determination.

ii) There are cases, like the one of the semileptonic form factor discussed in
the next section, in which the error due to the quenching of the strange quark
in the $N_f=2$ calculation is evaluated and included in the systematic
uncertainty. In other cases, the comparison between the $N_f=2$ and $N_f=2+1$
results shows, a posteriori, that the systematic effect due to the quenching of
the strange quark is not visible within the currently reached accuracy (to the
best of my knowledge, this is actually the case of all lattice calculations
performed so far). In all these cases, I do not see any valid reason for
ignoring the information coming from the $N_f=2$ calculations and for not
combining together the two sets of results.

iii) The $N_f=2+1$ theory is not really ``full QCD''. Indeed, $N_f=2+1+1$
lattice calculations are already being performed (see for
instance~\cite{Baron:2009zq}). I would find unreasonable to simply forget the
$N_f=2+1$ calculations when $N_f=2+1+1$ results will be available. For all those
quantities for which the error due to the quenching of the charm quark will turn
out (a posteriori) to be negligible, I will suggest again to average together
$N_f=2+1$ and $N_f=2+1+1$ results.

For these reasons, in deriving lattice averages for the present review, I'm not
going to follow the FLAG criterium as far as the number of dynamical flavours is
concerned, and I will rather address this issue on a case by case basis.

\section{$\mathbf{|V_{us}|}$ from semileptonic and leptonic kaon decays}

The most accurate determinations of the Cabibbo angle, or equivalently the CKM
matrix element $V_{us}$, come from the study of semileptonic $K \to \pi \ell
\nu$ ($K_{\ell 3}$) and leptonic $K \to \ell \nu$ ($K_{\ell 2}$) decays. Very
precise experimental measurements of the $K_{\ell 3}$ and of the ratio of
$K_{\ell 2}$ over $\pi_{\ell 2}$ decay rates allow to determine the following
combinations of CKM and hadronic parameters~\cite{flavianet},
\beq
\vert V_{us} \vert \, f_+(0) = 0.21664(48) \qquad , \qquad
\left| \frac{V_{us}}{V_{ud}} \right| \frac{f_K}{f_\pi} = 0.27599(59) \ ,
\label{eq:vus-exp}
\eeq
with an accuracy of about 2\textperthousand. In eq.~(\ref{eq:vus-exp}), $f_+(0)$
is by convention the form factor for the $K^0 \to \pi^{-}$ matrix element.
Moreover, both $f_+(0)$ and the ratio of decay constants $f_K/f_\pi$ in
eq.~(\ref{eq:vus-exp}) are defined in the isospin symmetric limit $m_u = m_d$
(keeping the kaon and pion masses to their physical value) and neglecting
electromagnetic corrections. These are therefore the quantities that are
directly determined in lattice QCD simulations.

A determination of the hadronic parameters $f_+(0)$ and $f_K/f_\pi$, which
assumes the validity of the Standard Model and it is independent of lattice QCD
calculations, has been provided by FLAG~\cite{flag}. It makes use of the first
row unitarity constraint on the CKM matrix,
\beq
\vert V_{ud} \vert^2 + \vert V_{us} \vert^2 + \vert V_{ub} \vert^2 =1 \ ,
\label{eq:unitarity}
\eeq
which holds in the Standard Model. Since within present uncertainties the
contribution of $\vert V_{ub} \vert$ in eq.~(\ref{eq:unitarity}) is numerically
negligible, the unitarity constraint~(\ref{eq:unitarity}) can be combined
with the two experimental results in (\ref{eq:vus-exp}) to provide a set of
three equations and four unknowns: $\vert V_{ud}\vert$, $\vert V_{us}\vert$,
$f_+(0)$ and $f_K/f_\pi$. A precise determination of $\vert V_{ud}\vert$ is
provided by the study of nuclear $\beta$-decays~\cite{Hardy:2008gy}, based on
the analysis of 20 different superallowed transitions:
\beq
\vert V_{ud} \vert  = 0.97425(22) \ .
\label{eq:vud}
\eeq
Using this result, one can then obtain a determination of the other three
parameters,
\beq
\vert V_{us} \vert  = \left( 1 -\vert V_{ud} \vert^2 \right)^{1/2} = 0.22547(95)
\label{eq:flag-vus}
\eeq
\beq
f_+(0) = 0.9608(46) \qquad , \qquad
\frac{f_K}{f_\pi} = 1.1925(56) \ .
\label{eq:flag-hpar}
\eeq
The estimates~(\ref{eq:flag-hpar}) of the hadronic parameters are benchmarks for
the lattice results reviewed in this talk. Since both $f_+(0)$ and $f_K/f_\pi$
are equal to 1 in the SU(3)-symmetric limit, what it is actually measured on the
lattice are the SU(3) breaking effects, i.e. the differences $f_+(0)-1$ and
$f_K/f_\pi-1$. Eq.~(\ref{eq:flag-hpar}) shows that, in order to provide a
significant test of the Standard Model, these differences must be determined on
the lattice with an accuracy of about 10\% and 3\% respectively. Of course, when
exploiting new physics scenarios beyond the Standard Model, the unitarity
relation (\ref{eq:unitarity}) should not be assumed and the determinations in
eq~(\ref{eq:flag-hpar}) are no longer valid.

\subsection{Semileptonic kaon decays: $\mathbf{f_+(0)}$}
The lattice determinations of $f_+(0)$ and, in the next section, of $f_K/f_\pi$,
are now reviewed.
 
The standard approach to study the vector form factor of $K_{\ell 3}$ decays is
based on SU(3) ChPT. In this framework, the vector form factor at zero momentum
transfer has an expansion of the form $f_+(0) = 1 + f_2 + f_4 + \ldots$, where
$f_n=O[m_{K,\pi}^n/(4\pi f_\pi)^n]$ and the first term of the expansion is equal
to 1 due to the current conservation in the SU(3)-symmetric limit. The
Ademollo-Gatto theorem~\cite{ag} shows that the deviation of $f_+(0)$ from 1 is
at least quadratic in the breaking of SU(3). Moreover, the first correction
$f_2$ receives contribution only from chiral loops and can be computed
unambiguously in terms of the kaon and pion masses and the pion decay constant.
It takes the value $f_2 = -0.0226$~\cite{gl}. The problem of estimating $f_+(0)$
can be thus re-expressed as the problem of finding a prediction for $\Delta f =
f_+(0) - (1 + f_2)$. The reference estimate for this quantity is still the one
obtained by Leutwyler and Roos in 1984~\cite{lr}, using a general
parameterization of the SU(3) breaking structure of the pseudoscalar meson wave
functions. It reads $\Delta f =-0.016(8)$, which implies $f_+(0) = 0.961(8)$.

Lattice QCD studies of $K_{\ell 3}$ decays started only relatively recently. The
strategy, which allows to reach the required percent accuracy in the
determination of the vector form factor, has been developed in
ref.~\cite{Becirevic:2004ya}, and it is based on the calculation of the scalar
form factor at maximum momentum transfer ($q^2_{max}=(m_K-m_\pi)^2$) through
the ratio
\beq
\frac{ \langle \pi |\bar s \gamma_0 u |K \rangle \langle \pi |\bar s \gamma_0 u
|K \rangle} {\langle \pi |\bar u \gamma_0 u |\pi \rangle  \langle K |\bar s
\gamma_0 s |K \rangle} = \frac{(m_K+m_\pi)^2}{4\, m_K m_\pi}
f_0(q^2_{max},\,m_K^2, \,m_\pi^2) \ ,
\label{eq:doubleratio}
\eeq
where the external states are of pion and kaon at rest. The double ratio on the
l.h.s. of eq.~(\ref{eq:doubleratio}) is equal to 1 in the SU(3) symmetric limit,
and it can be evaluated on the lattice with sub-percent accuracy at the
simulated values of pion and kaon masses. The physical form factor at zero
momentum transfer, $f_+(0)=f_0(0)$, is then obtained by extrapolating
$f_0(q^2_{max},\,m_K^2, \,m_\pi^2)$ to $q^2=0$ and to the physical meson masses.

A list of lattice results for $f_+(0)$ is collected in
table~\ref{tab:f+(0)-ris}, with the colour code in the FLAG style assigned
for each calculation. The relevant simulation parameters for these calculations
are given in table~\ref{tab:f+(0)-par}. 
\begin{table}[t]
\centering 
\vspace{2.0cm}{\footnotesize
\begin{tabular}{llclllll}
Collaboration & Ref. & $N_f$ &
\begin{rotate}{60}{publication status}\end{rotate}&
\begin{rotate}{60}{chiral extrapolation}\end{rotate}&
\begin{rotate}{60}{finite volume errors}\end{rotate}&
\begin{rotate}{60}{continuum extrapolation}\end{rotate}&
$f_+(0)$ \\
&&&&&& \\[-0.3cm] \hline \hline &&&&&& \\[-0.1cm]
\fcolorbox{red}{white}{RBC/UKQCD 07}   & \cite{Boyle:2007qe}  &2+1 
&\gA &\tbo&\tbg&\tbr & 0.9644(33)(34)(14)\\[1.0mm]
&&&&&& \\[-0.1cm] \hline &&&&&& \\[-0.1cm]
\fcolorbox{red}{white}{ETMC 09}  & \cite{Lubicz:2009ht}  &2  
&\gA &\tbo&\tbo&\tbo & 0.9560(57)(62)\\[1.0mm]
QCDSF 07       & \cite{Brommel:2007wn} &2  &\rC &\tbr&\tbg&\tbr &
0.9647(15)$_{stat}$\\[1.0mm]
RBC 06         & \cite{Dawson:2006qc}  &2  &\gA &\tbr&\tbg&\tbr &
0.968(9)(6)\\[1.0mm]
JLQCD 05       & \cite{Tsutsui:2005cj} &2  &\rC &\tbr&\tbg&\tbr &
0.967(6)\\[1.0mm]
&&&&&& \\[-0.1cm] \hline &&&&&& \\[-0.1cm]
SPQ$_{\mbox{\tiny CD}}$R 04 & \cite{Becirevic:2004ya} &0   &\gA &\tbr&\tbg&\tbr
& 0.960(5)(7)\\[1.0mm]
&&&&&& \\[-0.1cm] \hline \hline
\end{tabular}}
\caption{Colour code rating in the FLAG style and summary of lattice results for
$f_+(0)$. The results discussed at this conference are framed.
\label{tab:f+(0)-ris}}
\vspace{0.8cm}{\footnotesize
\begin{tabular}{llclccc}
Collaboration & Ref. & $N_f$ & action &$a$/fm& $\hspace{-0.3cm}(Lm_\pi)^{\rm
min}\hspace{-0.3cm}$ & $m_\pi$/MeV \\
&&&&& \\[-0.3cm] \hline \hline &&&&& \\[-0.1cm]
\fcolorbox{red}{white}{RBC/UKQCD 07} & \cite{Boyle:2007qe} &2+1 &DWF 
&0.11 & 4.6& $\gtrsim 330$ \\[1.0mm]
&&&&& \\[-0.1cm] \hline &&&&& \\[-0.1cm]
\fcolorbox{red}{white}{ETMC 09} & \cite{Lubicz:2009ht} & 2 &max. tmQCD 
&$\gtrsim 0.07$& 3.7& $\gtrsim 260$ \\[1.0mm]
QCDSF 07& \cite{Brommel:2007wn} & 2 & clover (NP) &0.08 & 5.4& $\gtrsim 590$
\\[1.0mm]
RBC 06  & \cite{Dawson:2006qc}  & 2 & DWF &0.12 & 4.7& $\gtrsim 490$ \\[1.0mm]
JLQCD 05& \cite{Tsutsui:2005cj} & 2 &clover (NP) &0.09 & 4.9& $\gtrsim 550$
\\[1.0mm]
&&&&& \\[-0.1cm] \hline &&&&& \\[-0.1cm]
SPQ$_{\mbox{\tiny CD}}$R 04 & \cite{Becirevic:2004ya}& 0 & clover (NP) &0.07 &
4.6& $\gtrsim 500$ \\[1.0mm]
&&&&& \\[-0.1cm] \hline \hline
\end{tabular}}
\caption{Parameters of the simulations listed in
table~\protect\ref{tab:f+(0)-ris}.
\label{tab:f+(0)-par}}
\end{table}
The first lattice calculation of the form factor, by the SPQ$_{\mbox{\tiny
CD}}$R collaboration~\cite{Becirevic:2004ya}, was performed in the quenched
approximation and with rather large values of simulated pion masses, $m_\pi
\gtrsim 500$ MeV. It gave the result $f_+(0) = 0.960(5)(7)$, in remarkable
agreement with the quark model prediction by Leutwyler and Roos~\cite{lr}.
Subsequently, unquenched
calculations~\cite{Brommel:2007wn,Dawson:2006qc,Tsutsui:2005cj}, performed with
$N_f=2$ dynamical fermions but still large values of simulated pion masses,
confirmed the quenched result. The first lattice calculation of $f_+(0)$ which
aims to keep under control all sources of systematic uncertainties has been
performed by the RBC/UKQCD collaboration~\cite{Boyle:2007qe}. This year, a
second calculation of comparable accuracy has been presented by
ETMC~\cite{Lubicz:2009ht}. Both the update of the RBC/UKQCD~07 result and
the new ETMC~09 calculation have been discussed in the parallel session at this
conference~\cite{jzanotti,DiVita:2009by}.

The RBC/UKQCD calculation~\cite{Boyle:2007qe} uses the DWF action with $N_f=2+1$
dynamical quarks, and pion masses as light as 330 MeV. The colour code rating
for this calculation, displayed in table~\ref{tab:f+(0)-ris}, includes a red
square for the continuum extrapolation since the simulation has been performed
at a single value of the lattice spacing ($a \simeq 0.11$ fm). A 4\% of
systematic error on $(1-f_+)$ due to discretization effects is assigned on the
basis of the parametric estimate $\sigma_{discr.} \sim (a \Lambda_{QCD})^2$,
assuming $\Lambda_{QCD} \simeq 300$ MeV. In the parallel talk at this
conference~\cite{jzanotti}, James Zanotti for RBC/UKQCD has announced that a
simulation at a second finer lattice spacing ($a \simeq 0.09$ fm) is being
performed and corresponding results for the vector form factor will be presented
soon. Two other improvements on the existing calculation are being implemented,
namely a simulation at a second value of the strange quark mass (the value of
$m_s$ firstly simulated by RBC/UKQCD~\cite{Boyle:2007qe} was about 15\% heavier
than the physical strange quark) and the use of twisted boundary conditions. The
latters, applied to valence quark
fields~\cite{Boyle:2007wg,Guadagnoli:2005be,Sachrajda:2004mi}, allow to simulate
very close or directly at $q^2=0$, thus removing the model dependence associated
with the momentum extrapolation.

The ETMC calculation~\cite{Lubicz:2009ht} of $f_+(0)$ has been presented at this
conference by Silvano Si\-mu\-la~\cite{DiVita:2009by}. It is based on the set of
$N_f=2$ simulations performed by ETMC with twisted mass fermions at maximal
twist. Finite size effects on the form factor have been estimated by simulating
on two different volumes, whereas discretization errors have been evaluated by
performing, for a single value of the light quark mass, calculations at three
different lattice spacing ($a \simeq 0.07,\, 0.09,\,0.10$ fm).\footnote{The
results on the coarsest lattice have been analysed after the conference and
are discussed in the proceedings~\cite{DiVita:2009by}.} In order to get results
close to $q^2=0$, twisted boundary conditions on the valence quark fields have
been implemented. Both pole and quadratic fits have been considered to
interpolate to $q^2=0$. In the main simulation on the lattice with $a\simeq
0.09$ fm, 6 different values of the light quark masses have been simulated, with
the lightest pion mass being $m_\pi \simeq 260$ MeV.

The relatively large number of light quark masses simulated by ETMC allows to
achieve a good control over the chiral extrapolation, which represents one of
the main source of systematic uncertainty in the lattice evaluation of $f_+(0)$.
The chiral extrapolation has been performed using both SU(3) and, for the first
time, SU(2) ChPT. The use of SU(2) ChPT applied to kaon
ob\-ser\-va\-bles~\cite{Roessl:1999iu} has been mainly advocated, in the context
of lattice calculations, by RBC/UKQCD~\cite{Allton:2008pn}. For a general
discussion on the applicability of the different versions of ChPT to lattice
results see the plenary talk by Enno Scholz at this
conference~\cite{Scholz:2009yz}. For the vector form factor $f_+(0)$, the SU(2)
chiral expansion has been derived at the NLO by Flynn and
Sachrajda~\cite{Flynn:2008tg}. ETMC finds that the convergence of the SU(2)
expansion for $f_+(0)$ is indeed very good, and contributions beyond NLO are
small both at the physical point and in the region of simulated lattice data.
Instead, both ETMC and RBC/UKQCD observe that the contribution beyond the NLO in
the SU(3) chiral expansion, i.e. the quantity $\Delta f = f_+(0) - (1 + f_2)$,
is large and of the same size of the NLO contribution $f_2$. Nevertheless, the
analyses based on SU(2) and SU(3) ChPT performed by ETMC leads eventually to
consistent results, and their difference is included in the final estimate of
the systematic error.

The ETMC calculation is performed with $N_f=2$ dynamical flavours. The effects
due to the strange quark loops are exactly accounted for, in the calculation, up
to the NLO in the SU(3) chiral expansion. This has been possible, because the
NLO term in the chiral expansion, $f_2$, can be precisely computed (in terms of
pion and kaon masses only) for the theories with
$N_f=0$~\cite{Becirevic:2004ya}, $N_f=2$~\cite{Becirevic:2005py} and
$N_f=2+1$~\cite{gl} dynamical quarks. Thus, the only uncertainty due to the
quenching of the strange quark in the ETMC calculation concerns the NNLO
contribution $\Delta f$. For this quantity, the relative uncertainty has been
estimated by ETMC to be of the order of 13\% which turns out to be of the same
size of the entire difference $(\Delta f)^{Nf=2}-(\Delta f)^{Nf=0}$, evaluated
using the quenched estimate of $\Delta f$ of ref.~\cite{Becirevic:2004ya}. This
difference is most likely an overestimate of the true error affecting $\Delta f$
in the $N_f=2$ calculation, i.e. $(\Delta f)^{phys.}-(\Delta f)^{Nf=2}$.

In order to quote a lattice average for $f_+(0)$, I will take into account both
the RBC/UKQCD and ETMC results, i.e.
\beq
\renewcommand{\arraystretch}{1.2}
\begin{array}{ll}
f_+(0) = 0.964(3)(4) \quad & [\ N_f=2+1 \ , \ {\rm RBC/UKQCD~07}\ ] \\
f_+(0) = 0.956(6)(6) \quad & [\ N_f=2 \ , \ {\rm ETMC~09}\ ] \ .
\end{array}
\renewcommand{\arraystretch}{1.0}
\label{eq:fp0-2results}
\eeq
These are the only results which are obtained with relatively light pion masses
and do not get therefore in table~\ref{tab:f+(0)-ris} a red square for the
chiral extrapolation. The RBC/UKQCD calculation has a red square assigned for
the continuum extrapolation, but this source of error is subdominant in the
calculation (even if the estimated error of 4\% were doubled, the impact on the
final result would be small). On the other hand, except for the partial
quenching, the $N_f=2$ calculation by ETMC has at present a better control over
the other systematic uncertainties with respect to the $N_f=2+1$ calculation by
RBC/UKQCD (3 lattice spacings rather than 1, 6 pion masses rather than 4, both
SU(3) and SU(2) chiral extrapolations). Since the uncertainty due to the
quenching of the strange quark has been accounted for in the ETMC result, I
quote as the best estimate of $f_+(0)$ the average of the two results
in eq.~(\ref{eq:fp0-2results}), obtaining
\beq
\setlength{\fboxrule}{2pt}
\fcolorbox{orange}{white}{$f_+(0) = 0.962(3)(4)$} \ .
\label{eq:fp0final}
\eeq
The first error in eq.~(\ref{eq:fp0final}) is statistical, and it is evaluated
in the standard way assuming gaussian statistical uncertainties. The second
error is systematic, and it has been taken to be equal to the systematic error
quoted by RBC/UKQCD.\footnote{While this contribution was almost finished to be
written, an updated estimate of the form factor has been presented by
RBC/UKQCD~\cite{Boyle:2010bh}, which reads
\beq
f_+(0) = 0.9599(34)(^{+31}_{-43})(14)
\label{eq:fp0rbcukqcd10} ~.
\eeq
The new analysis includes the simulation at a second value of the strange quark
mass and twisted boundary conditions to simulate directly at $q^2=0$, as
anticipated at this conference.  With respect to ref.~\cite{Boyle:2007qe}, the
change in the central value, as well as the second error in
eq.~(\ref{eq:fp0rbcukqcd10}), are due to the uncertainty in the chiral
extrapolation, which was not considered in ref.~\cite{Boyle:2007qe}. It has been
evaluated by varying in the SU(3) chiral expansion the value of the LO
low-energy constant $f$ in the range 100-131 MeV (ETMC finds that the effect of
a similar change in its chiral fit is smaller than the error already assigned
to the chiral extrapolation~\cite{simula}). Since the updated
result~(\ref{eq:fp0rbcukqcd10}) is in very good agreement with the lattice
average given in eq.~(\ref{eq:fp0final}), I find unnecessary to change this
average.}

The lattice average~(\ref{eq:fp0final}) is in very good agreement with the
lattice-independent estimate quoted in eq.~(\ref{eq:flag-hpar}), based on CKM
unitarity and the determination of $V_{ud}$ from nuclear $\beta$-decays.
Combining the lattice determination~(\ref{eq:fp0final}) with the experimental
result of eq.~(\ref{eq:vus-exp}), one obtains the estimate
\beq
\vert V_{us} \vert _{K_{\ell 3}} = 0.2252(13) \ ,
\label{eq:vus_kl3}
\eeq
in good agreement with the unitarity determination in eq.~(\ref{eq:flag-vus}).

\begin{figure}
\begin{center}
\includegraphics[width=.6\textwidth]{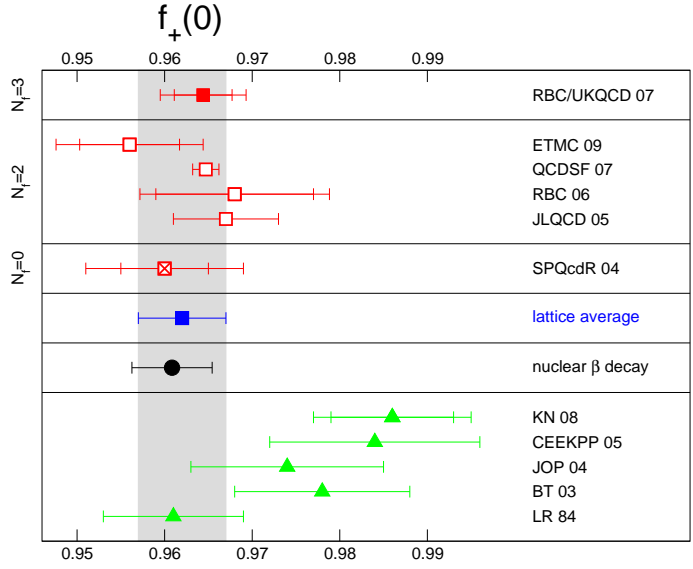}
\vspace{-0.2cm}
\end{center}
\caption{{\sl Summary of results for the vector form factor $f_+(0)$ of kaon
semileptonic decays. Lattice results, as obtained from simulations with
$N_f=0,2,3$ dynamical quarks, are shown in the upper side of the plot by red
squares. The results of analytical model
calculations~\cite{lr},\cite{Kastner:2008ch}-\cite{Bijnens:2003uy} are shown in
the lower side of the plot by green triangles. The black circle represents the
lattice-independent estimate of $f_+(0)$ given in
eq.~(\protect\ref{eq:flag-hpar}), which is only valid in the Standard Model. The
blue square and the grey band show the average of lattice results for $f_+(0)$
derived in eq.~(\protect\ref{eq:fp0final}). This plot has been produced by FLAG
and updated for the lattice conference.}}
\label{fig:f+(0)}
\vspace{-0.2cm}
\end{figure}

A summary of lattice results for $f_+(0)$ is shown in fig.~\ref{fig:f+(0)},
which is an updated version for the lattice conference of a plot produced by
FLAG~\cite{flag}.
The blue square and grey band in the plot represent the lattice
average~(\ref{eq:fp0final}). The lattice-independent
estimate~(\ref{eq:flag-hpar}) of $f_+(0)$, obtained using the unitarity
determination of $V_{us}$, is also shown for comparison. In the lower side of
the plot, we collect the quark model result of Leutwyler and Roos
(LR~84)~\cite{lr} together with the results of more recent analytical model
calculations~\cite{Kastner:2008ch}-\cite{Bijnens:2003uy}. These latter
determinations of $f_+(0)$ turn out to be larger than both the value predicted
by lattice QCD and the result implied by CKM unitarity.

\subsection{Leptonic kaon decays: $\mathbf{f_K/f_\pi}$}
A detailed review of lattice results for the pion and kaon decay constants has
been given by Enno Scholz in his plenary talk at this
conference~\cite{Scholz:2009yz}. For this reason, in this talk I will only
summarize the main features of the new calculations and derive a lattice average
for the ratio $f_K/f_\pi$, which is the relevant hadronic input parameter for
the determination of the Cabibbo angle.

A list of unquenched lattice results for $f_K/f_\pi$ is collected in
table~\ref{tab:FKFpi-ris} (based on the FLAG work), with the colour code rating
in  the FLAG style assigned for each calculation. The relevant simulation
parameters for these calculations are given in table~\ref{tab:FKFpi-par}.

A first look at tables~\ref{tab:FKFpi-ris} and \ref{tab:FKFpi-par} is sufficient
to reveal the large number of unquenched predictions for $f_K/f_\pi$ obtained by
the various lattice collaborations in the last one or two years. For the purpose
of this conference, the results which are new or have been updated in the last
year are framed in the tables. Moreover, several lattice predictions for
$f_K/f_\pi$ in table~\ref{tab:FKFpi-ris} are rated without red squares,
indicating that all sources of systematic uncertainties in these calculations
are kept sufficiently well under control.
\begin{table}[p]
\centering 
\vspace{1.7cm}{\footnotesize
\begin{tabular}{llclllll}
Collaboration & Ref. & $N_f$ &\hspace{0.06cm}
\begin{rotate}{60}{publication status}\end{rotate}&\hspace{0.06cm}
\begin{rotate}{60}{chiral extrapolation}\end{rotate}&\hspace{0.06cm}
\begin{rotate}{60}{finite volume errors}\end{rotate}&\hspace{0.06cm}
\begin{rotate}{60}{continuum extrapolation}\end{rotate}&\hspace{0.06cm}
$f_K/f_\pi$ \\
&&&&&& \\[-0.3cm] \hline \hline &&&&&& \\[-0.3cm]
\fcolorbox{red}{white}{ALVdW 09} &\cite{ALVdW09} &2+1 &\rC &\tbo&\tbo&\tbo &
1.192(12)(16)\\[1.0mm]
\fcolorbox{red}{white}{BMW 09} &\cite{Ramos:2010ar,Durr:2010hr}  &2+1 &\oP
&\tbg&\tbg&\tbg & 1.192(7)(6)\\[1.0mm]
\fcolorbox{red}{white}{RBC/UKQCD 09} &\cite{Mawhinney:2009jy} &2+1 &\rC
&\tbo&\tbg&\tbo & 1.225(12)(14)\\[1.0mm]
\fcolorbox{red}{white}{MILC 09b} &\cite{Bazavov:2009tw} &2+1 &\gA
&\tbg&\tbg&\tbg & 1.198(2)($^{\;+6}_{-8}$)\\[1.0mm]
MILC 09a   &\cite{Bazavov:2009bb}  &2+1 &\gA &\tbg&\tbg&\tbg &
1.197(3)($^{\;+6}_{-13}$)\\[1.0mm]
\fcolorbox{red}{white}{JLQCD/TWQCD 09} &\cite{JLQCD:2009sk} &2+1 &\rC
&\tbo&\tbr&\tbr & 1.210(12)$_{stat}$\\[1.0mm]
PACS-CS 08 &\cite{Aoki:2008sm} &2+1 &\gA &\tbg&\tbr&\tbr & 1.189(20)\\[1.0mm]
HPQCD/UKQCD 07 &\cite{Follana:2007uv} &2+1 &\gA &\tbg&\tbo&\tbg& 
1.189(2)(7) \\ [1.0mm]
RBC/UKQCD 08  &\cite{Allton:2008pn} &2+1 &\gA &\tbo&\tbg&\tbr &
1.205(18)(62)\\[1.0mm]
NPLQCD 06  &\cite{Beane:2006kx} &2+1 &\gA &\tbo&\tbr&\tbr &
1.218(2)($^{+11}_{-24}$)\\[1.0mm]
MILC 04 &\cite{Aubin:2004fs} &2+1 &\gA &\tbg&\tbo&\tbo & 1.210(4)(13) \\
&&&&&& \\[-0.3cm] \hline &&&&&& \\[-0.3cm]
\fcolorbox{red}{white}{ETMC 09} &\cite{Blossier:2009bx} &2 &\gA &\tbo&\tbo&\tbg
& 1.210(6)(15)(9)\\[1.0mm]
ETMC 07 &\cite{Blossier:2007vv} &2 &\gA &\tbo&\tbo&\tbr & 1.227(9)(24)\\[1.0mm]
QCDSF/UKQCD 07 &\cite{QCDSF07}  &2 &\rC &\tbo&\tbg&\tbo & 1.21(3)\\
&&&&&& \\[-0.3cm] \hline \hline &&&&&& \\[-0.3cm]
\end{tabular}}
\caption{Colour code rating in the FLAG style and summary of unquenched lattice
results for $f_K/f_\pi$. The new results, obtained in the last year, are framed.
\label{tab:FKFpi-ris}}
\vspace{1.0cm}
{\footnotesize
\begin{tabular}{llclccc}
Collaboration & Ref. & $N_f$ & action & $a$/fm & \hspace{-0.3cm}$(Lm_\pi)^{\rm
min}$\hspace{-0.3cm} & $m_\pi$/MeV \\
&&&&& \\[-0.3cm] \hline \hline &&&&& \\[-0.3cm]
\fcolorbox{red}{white}{ALVdW 09} &\cite{ALVdW09} &2+1 & ${\rm KS_{\rm
MILC}/{\rm DWF}}$ & $\gtrsim 0.09$ & 3.8 & $\gtrsim300$ \\[1.0mm]
\fcolorbox{red}{white}{BMW 09} &\cite{Ramos:2010ar,Durr:2010hr} &2+1 & impr.
Wilson & $\gtrsim 0.07$ & 4.0 & $\gtrsim190$ \\[1.0mm]
\fcolorbox{red}{white}{RBC/UKQCD 09} &\cite{Mawhinney:2009jy} &2+1 & DWF &
$\gtrsim 0.08$ & 4.0 & $\gtrsim290$ \\[1.0mm]
\fcolorbox{red}{white}{MILC 09b} &\cite{Bazavov:2009tw} &2+1   & ${\rm KS_{\rm
MILC}^{\rm MILC}}$ & $\gtrsim 0.045$ & 4.0 & $\gtrsim230$ \\[1.0mm]
MILC 09a   &\cite{Bazavov:2009bb} &2+1 & ${\rm KS_{\rm MILC}^{\rm MILC}}$ &
$\gtrsim 0.045$ & 3.8 & $\gtrsim230$ \\[1.0mm]
\fcolorbox{red}{white}{JLQCD/TWQCD 09} &\cite{JLQCD:2009sk} &2+1 & Overlap &
0.10 & 2.8 & $\gtrsim 340$ \\[1.0mm]
PACS-CS 08 &\cite{Aoki:2008sm} &2+1 & clover (NP) & 0.09 & 2.3 & $\gtrsim160$ 
\\[1.0mm]
HPQCD/UKQCD 07 &\cite{Follana:2007uv} &2+1 &$\rm KS^{\rm HISQ}_{\rm MILC}$ &
$\gtrsim 0.09$ & 3.8 & $\gtrsim250$ \\ [1.0mm]
RBC/UKQCD 08   &\cite{Allton:2008pn} &2+1 & DWF & 0.11 & 4.6 & $\gtrsim330$ 
\\[1.0mm]
NPLQCD 06  &\cite{Beane:2006kx}  &2+1 & $\rm KS_{\rm MILC}/DWF$ & 0.13 & 3.7 &
$\gtrsim290$ \\[1.0mm]
MILC 04 &\cite{Aubin:2004fs} &2+1 & ${\rm KS_{\rm MILC}^{\rm MILC}}$ & $\gtrsim
0.09$ & 3.8 & $\gtrsim250$ \\
&&&&& \\[-0.3cm] \hline &&&&& \\[-0.3cm]
\fcolorbox{red}{white}{ETMC 09} &\cite{Blossier:2009bx} &2  &max. tmQCD &
$\gtrsim 0.07$ & 3.2 & $\gtrsim260$ \\[1.0mm]
ETMC 07  &\cite{Blossier:2007vv} &2 & max. tmQCD  & 0.09  & 3.2 & $\gtrsim300$
\\[1.0mm]
QCDSF/UKQCD 07 &\cite{QCDSF07} &2 & clover (NP) & $\gtrsim 0.06$ & 4.2 &
$\gtrsim300$ \\
&&&&& \\[-0.3cm] \hline \hline &&&&& \\[-0.3cm]
\end{tabular}}
\caption{Parameters of the simulations listed in
table~\protect\ref{tab:FKFpi-ris}.
\label{tab:FKFpi-par}}
\end{table}

At this conference, an update of the MILC analysis has been presented by Urs
Heller~\cite{Bazavov:2009tw}. This analysis is based on SU(3) (rooted) staggered
ChPT and includes, as a new feature, the N$^2$LO continuum chiral logs. The
chiral fits are based on results obtained with the ``fine'', ``super-fine'' and
``ultra-fine'' MILC ensembles and are performed in two stages. The first one
only includes the lowest quark masses, and it is used to determine the LO and
NLO low energy constants. Once these constants are fixed, then the higher order
contributions, namely the complete N$^2$LO contribution together with N$^3$LO
and N$^4$LO analytic terms, are determined through a global fit over all quark
masses.

The RBC-UKQCD analysis of $f_K/f_\pi$, presented at this conference by Bob
Mawhinney~\cite{Mawhinney:2009jy}, is based on SU(2) chiral fits. While previous
results by the collaboration were obtained at a single value of the lattice
spacing ($a^{-1}=1.72$ GeV), the new analysis includes data from a second
ensemble with a finer lattice ($a^{-1}=2.32$ GeV), which also includes lighter
dynamical quarks. In addition, on the coarse lattice more configurations have
been generated, by almost doubling the statistics from earlier works.

An accurate prediction for $f_K/f_\pi$ has been also presented by the BMW
collaboration~\cite{Ramos:2010ar,Durr:2010hr}, based on their extensive
simulation performed at three values of the lattice spacing, large volumes and
simulated pion masses reaching down about 190 MeV. The chiral extrapolation is
performed testing three different assumptions for the quark mass dependence:
ChPT, either SU(3) or SU(2), or a simple polynomial extrapolation. With respect
to the preliminary determination presented in a poster by Alberto Ramos at this
conference, the final result for $f_K/f_\pi$ given
in~\cite{Ramos:2010ar,Durr:2010hr} and quoted in table~\ref{tab:FKFpi-ris} has a
reduced systematic uncertainty.

The JLQCD/TWQCD collaboration has presented a preliminary determination of the
pion and kaon decay constants~\cite{JLQCD:2009sk} based on a simulation with
$N_f=2+1$ dynamical overlap fermions, at five different up and down quark
masses. While the chosen approach is definitively a benchmark for lattice QCD,
since the overlap formulation preserves an exact chiral symmetry at finite
lattice spacing, the simulation performed by JLQCD/TWQCD relies yet on a single
value of the lattice spacing ($a\simeq 0.10$ fm) and a rather small lattice size
($(Lm_\pi)^{\rm min}<3$), two features which imply the red squares in the FLAG
stye rating of table~\ref{tab:FKFpi-ris}.

Among the other $N_f=2+1$ determinations of $f_K/f_\pi$ collected in
table~\ref{tab:FKFpi-ris} and not presented at this conference, the ALVdW 09 and
HPQCD/UKQCD 08 results are also free of red tags and will be thus included in
the determination of the final lattice average.

A determination of $f_K/f_\pi$ with $N_f=2$ dynamical quarks, in which all
sources of systematic uncertainties are kept under control (i.e. no red squares
in the colour code rating) has been recently obtained by the ETM
collaboration~\cite{Blossier:2009bx}, using maximally twisted mass fermions. The
comparison with the most precise $N_f=2+1$ calculations suggests that the
quenching effect of the strange quark in the determination of $f_K/f_\pi$ is
currently negligible, within the reached lattice accuracy. Therefore, I will
include this result in the determination of the final lattice average (in this
case, given the large number of independent determinations of $f_K/f_\pi$ which
are free of red squares in table~\ref{tab:FKFpi-ris}, the inclusion of the
ETM~09 result has a marginal impact). I then obtain
\beq
\setlength{\fboxrule}{2pt}
\fcolorbox{orange}{white}{$f_K/f_\pi = 1.196(1)(10)$} \ ,
\label{eq:fkfpifinal}
\eeq
where the first error is statistical, evaluated assuming gaussian statistical
uncertainties, and the second error is systematic. For the latter, I'm quoting
an uncertainty which is of the same size of the typical systematic uncertainty
estimated for the most precise lattice determinations listed in
table~\ref{tab:FKFpi-ris}.

The average~(\ref{eq:fkfpifinal}) can be translated into a determination of the
kaon decay constant, $f_K$, by using $f_\pi=130.41(3)(20)$
MeV~\cite{Rosner:2010ak}, as determined from the measurement of the leptonic
pion decay rate and $V_{ud}$ from eq.~(\ref{eq:vud}). In this way one gets
\beq
f_K = 156.0 \pm 1.3 ~ \mev \ .
\label{eq:fkfinal}
\eeq

The lattice average~(\ref{eq:fkfpifinal}) is also in very good agreement with
the lattice-independent estimate of $f_K/f_\pi$ quoted in
eq.~(\ref{eq:flag-hpar}), based on CKM unitarity and the determination of
$V_{ud}$ from nuclear $\beta$-decays. By combining the lattice
prediction~(\ref{eq:fkfpifinal}) with the experimental result of
eq.~(\ref{eq:vus-exp}) and with $V_{ud}$ from nuclear $\beta$-decays, one
obtains the estimate
\beq
\vert V_{us} \vert _{K_{\ell 2}} = 0.2249(19) \ .
\eeq
This in good agreement with both the determination~(\ref{eq:vus_kl3}) from
$K_{\ell 3}$ decays and with the unitarity determination in
eq.~(\ref{eq:flag-vus}).

A summary of lattice results for $f_K/f_\pi$ is shown in fig.~\ref{fig:f+(0)},
which is an updated version for the lattice conference of a plot produced by
FLAG~\cite{flag}.
\begin{figure}
\begin{center}
\includegraphics[width=.6\textwidth]{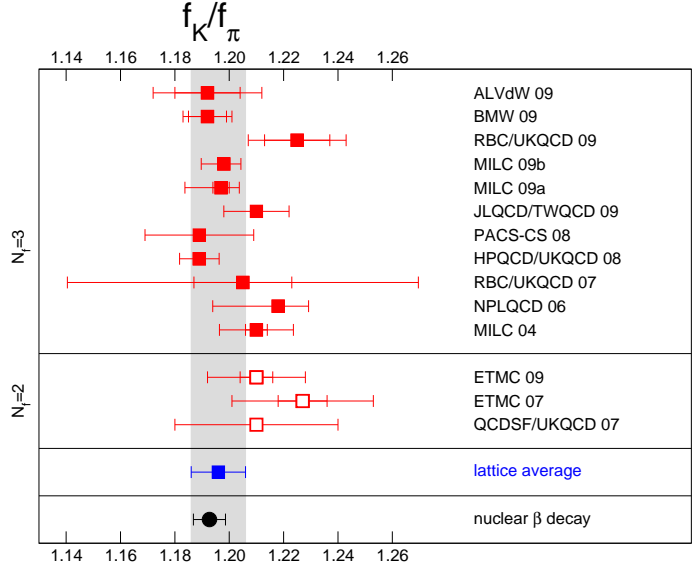}
\end{center}
\caption{{\sl Summary of unquenched lattice results for the ratio of decay
constants $f_K/f_\pi$ as obtained from simulations with $N_f=2$ and $N_f=3$
dynamical quarks (red squares). The black circle represents the
lattice-independent estimate of $f_K/f_\pi$ given in
eq.~(\protect\ref{eq:flag-hpar}), which is valid only in the Standard Model. The
blue square and the grey band show the average of the lattice results for
$f_K/f_\pi$ derived in eq.~(\protect\ref{eq:fkfpifinal}). This plot has been
produced by FLAG and updated for the lattice conference.}}
\label{fig:fkfpi}
\end{figure}
The blue square and grey band in the plot represents the lattice
average~(\ref{eq:fkfpifinal}). The lattice-independent
estimate~(\ref{eq:flag-hpar}) of $f_K/f_\pi$ is also shown for comparison.

\section{$\mathbf{K^0-\bar{K}^0}$ mixing: $\mathbf{B_K}$}

The accuracy in the lattice determination of the kaon bag parameter, $B_K$,
has remarkably improved over the last few years. This progress is well
illustrated by the following sample of lattice averages presented at the lattice
conferences:
\bea
\label{eq:bk-sharpe96}
{\tt Lattice~'96} \qquad & \hat{B}_K = 0.90 \pm 0.03 \pm 0.15 \qquad &
{\rm S.~Sharpe,~\cite{Sharpe:1996ih}} \\
\label{eq:bk-lellouch00}
{\tt Lattice~'00} \qquad & \hat{B}_K = 0.86 \pm 0.06 \pm 0.14 \qquad &
{\rm L.~Lellouch,~\cite{Lellouch:2000bm}} \\
\label{eq:bk-dawson95}
{\tt Lattice~'05} \qquad & \hat{B}_K = 0.79 \pm 0.04 \pm 0.08 \qquad &
{\rm C.~Dawson,~\cite{Dawson:2005za}} \\
\label{eq:bk-lellouch08}
{\tt Lattice~'08} \qquad & \hat{B}_K = 0.723 \pm 0.037 \qquad &
{\rm L.~Lellouch,~\cite{Lellouch:2009fg}}~ ,
\eea
where $\hat B_K$ is the renormalization group invariant definition (for $N_f=3$)
of the bag parameter. The second error quoted in
eqs.~(\ref{eq:bk-sharpe96})-(\ref{eq:bk-dawson95}) was an estimate of the
quenched error. This uncertainty has limited the accuracy of the lattice
calculations of $B_K$ for a long time. It has started to decrease when the first
unquenched estimates have been performed in the last few years, and in the
average of $B_K$ quoted at the last two lattice conferences it was definitively
removed. Nevertheless, all unquenched determinations of $B_K$ available until
last year were obtained at a fixed (and rather large) lattice spacing. Thus, a
quantitative estimate of discretization effects affecting these calculations,
which could have been not negligible according to the experience gathered in the
quenched approximation, was not available yet.

The list of unquenched lattice results for $B_K$ available today is collected in
table~\ref{tab:BK-ris}, with the colour code rating in the FLAG style assigned
for each calculation. The relevant simulation parameters for these calculations
are given in table~\ref{tab:BK-par}.
\begin{table}[!t]
\centering 
\vspace{2.0cm}{\footnotesize
\begin{tabular}{llcllllllll}
Collaboration  & Ref. & $N_f$ &
\begin{rotate}{60}{publication status}\end{rotate} &
\begin{rotate}{60}{continuum extrapolation}\end{rotate} &
\begin{rotate}{60}{chiral extrapolation}\end{rotate}&
\begin{rotate}{60}{finite volume errors}\end{rotate}&
\begin{rotate}{60}{renormalization}\end{rotate}  &
\begin{rotate}{60}{running}\end{rotate} &
$B_{\rm K}^{\msbar}(2\,\GeV)$ & $\hat{B}_{\rm K}$ \\
&&&&&&&& \\[-0.3cm] \hline \hline &&&&&&&& \\[-0.1cm]
\fcolorbox{red}{white}{ALVdW 09} & \cite{Aubin:2009jh} &2+1  &\gA & \tbo & \tbg
& \tbo & \tbg & \tbo & 0.527(6)(20)& 0.724(8)(28)\\[1.0mm]
\fcolorbox{red}{white}{RBC/UKQCD 09} & \cite{Kelly:2009fp} & 2+1 &\rC & \tbo &
\tbo & \tbg & \tbg & \tbo & 0.537(6)(18) & 0.738(8)(25) \\[1.0mm]
\fcolorbox{red}{white}{SBW 09} & \cite{Bae:2009tf}-\cite{Kim:2009ti} &2+1 
&\rC & \tbg & \tbg & \tbr & \tbr & \tbo & 0.512(14)(34)& 0.701(19)(47)\\[1.0mm]
RBC/UKQCD 07  & \cite{Antonio:2007pb,Allton:2008pn}  & 2+1 &\gA & \tbr & \tbo &
\tbg & \tbg & \tbo & 0.524(10)(28) & 0.720(13)(37) \\[1.0mm]
HPQCD/UKQCD 06  & \cite{Gamiz:2006sq} & 2+1 &\gA & \tbr & \tbo & \tbg & \tbr &
\tbo & 0.618(18)(135)& 0.83(18) \\[1.0mm]
&&&&&&&& \\[-0.1cm] \hline &&&&&&&& \\[-0.1cm]
\fcolorbox{red}{white}{ETMC 09} & \cite{Bertone:2009bu} & 2 &\rC & \tbg & \tbo &
\tbo & \tbg & \tbo & 0.518(21)(21) & 0.730(30)(30) \\[1.0mm]
JLQCD 08  & \cite{Aoki:2008ss}  & 2     &\gA & \tbr & \tbo & \tbr           &
\tbg & \tbo & 0.537(4)(40) & 0.758(6)(71)\\[1.0mm]
RBC 04    & \cite{Aoki:2004ht}  & 2     &\gA & \tbr & \tbr & \tbr$^\dagger$ &
\tbg & \tbo & 0.495(18)    & 0.699(25)   \\[1.0mm]
UKQCD 04   & \cite{Flynn:2004au} & 2     &\gA & \tbr & \tbr & \tbr$^\dagger$ &
\tbr & \tbo & 0.49(13)    & 0.69(18)    \\[1.0mm]
&&&&&&&& \\[-0.1cm]\hline \hline
\end{tabular}}
\caption{Colour code rating in the FLAG style and summary of unquenched lattice
results for $B_K$. The symbol \tbr$^\dagger$ means that these results have been
obtained at $( M_\pi L)_{\rm min} > 4$ in a lattice box with a spatial extension
$L < 2$~fm. The new results obtained in the last year are framed.
\label{tab:BK-ris}}
\vspace{1.0cm}{\footnotesize
\begin{tabular}{llcccccl}
Collaboration  & Ref. & $N_f$ & action & $a$/fm & \hspace{-0.3cm}$(Lm_\pi)^{\rm
min}$\hspace{-0.3cm} & $m_\pi$/MeV & Ren. \\
&&&&& \\[-0.3cm] \hline \hline &&&&& \\[-0.1cm]
\fcolorbox{red}{white}{ALVdW 09} & \cite{Aubin:2009jh} &2+1 & ${\rm KS_{\rm
MILC}/{\rm DWF}}$ & $\gtrsim 0.09$ & 3.5 & $\gtrsim230$ & RI \\[1.0mm]
\fcolorbox{red}{white}{RBC/UKQCD 09} & \cite{Kelly:2009fp} & 2+1 & DWF &
$\gtrsim 0.08$ & 4.0 & $\gtrsim290$ & RI \\[1.0mm]
\fcolorbox{red}{white}{SBW 09} & \cite{Bae:2009tf}-\cite{Kim:2009ti} &2+1 &
${\rm KS_{\rm MILC}^{\rm HYP}}$ & $\gtrsim 0.06$ & 2.5 & $\gtrsim 200$ &
PT1$\ell$ \\[1.0mm]
RBC/UKQCD 07 & \cite{Antonio:2007pb,Allton:2008pn} & 2+1 & DWF &0.11& 4.6&
$\gtrsim330$& RI \\[1.0mm]
HPQCD/UKQCD 06 & \cite{Gamiz:2006sq} & 2+1 & ${\rm KS_{\rm MILC}^{\rm MILC}}$
&0.12& 4.6& $\gtrsim360$& PT1$\ell$ \\[1.0mm]
&&&&& \\[-0.1cm] \hline &&&&& \\[-0.1cm]
\fcolorbox{red}{white}{ETMC 09} & \cite{Bertone:2009bu} & 2 & tmQCD/OS &
$\gtrsim 0.07$& 3.2& $\gtrsim260$ &RI \\[1.0mm]
JLQCD 08 & \cite{Aoki:2008ss} & 2 & overlap &0.12& 2.8& $\gtrsim290$ &RI
\\[1.0mm]
RBC 04 & \cite{Aoki:2004ht} & 2 & DWF &0.12& 4.6& $\gtrsim490$ &RI \\[1.0mm]
UKQCD 04 & \cite{Flynn:2004au}& 2 & clover(NP) &0.10& 6.2& $\gtrsim740$
&PT1$\ell$ \\[1.0mm]
&&&&& \\[-0.1cm] \hline \hline 
\end{tabular}}
\caption{Parameters of the simulations listed in table~\protect\ref{tab:BK-ris}.
\label{tab:BK-par}}
\end{table}

The novelty this year are a number of new calculations, namely
ALVdW~09~\cite{Aubin:2009jh}, RBC/ UKQCD~09~\cite{Kelly:2009fp},
SBW~09~\cite{Bae:2009tf}-\cite{Kim:2009ti} and ETMC~09~\cite{Bertone:2009bu},
which are framed in table~\ref{tab:BK-ris} for better illustration. At variance
with the previous unquenched calculations, they all involve an extrapolation to
the continuum limit for $B_K$, based on two (ALVdW~09, RBC/UKQCD~09) or three
(SBW~09, ETMC~09) values of the lattice spacing.

While the RBC/UKQCD~09, SBW~09 and ETMC~09 results are still preliminary, having
been only presented in the proceedings of this conference, the ALVdW~09
calculation is already published~\cite{Aubin:2009jh}. It uses a mixed action
setup, with domain wall fermions valence quarks over the $N_f=2+1$ staggered
gauge field configurations produced by MILC (``fine'' and ``coarse'' lattices).
The choice of domain wall fermions for the valence allows a straightforward
implementation of the RI-MOM method in order to non-perturbatively renormalize
the four-fermion operator relevant for $B_K$ (in the previous determination of
$B_K$ with staggered fermions by HPQCD/UKQCD~\cite{Gamiz:2006sq}, the one-loop
perturbative determination of the renormalization constant turned out to be
affected by an uncomfortably large systematic uncertainty). With domain wall
fermions, due to the residual breaking of chiral symmetry which is allowed by
the finite extension of the lattice in the fifth dimension, the $B_K$ operator
has a small mixing with operators of wrong chirality (but not with those of
incorrect taste). The final result for the bag parameter is obtained after
performing a combined chiral and continuum extrapolation based on NLO SU(3)
``mixed action'' ChPT~\cite{Bar:2005tu,Aubin:2006hg}, with the inclusion of some
analytic N$^2$LO contribution. The final accuracy quoted for $B_K$ is about 4\%,
where the dominant source of uncertainty is represented by the determination of
the renormalization constant.

The new result for $B_K$ obtained by RBC/UKQCD, which updates the first precise
unquenched calculation of ref.~\cite{Antonio:2007pb,Allton:2008pn}, has been
presented by Chris Kelly at this conference~\cite{Kelly:2009fp}. The main
improvement with respect to the previous calculation is the use of the second
finer lattice ensemble simulated by RBC/UKQCD, which allows to perform a
combined SU(2)-chiral and continuum extrapolation of $B_K$. This, in turn, has
permitted a substantial reduction of the systematic error (~3.5\%), since in the
previous calculation discretization effects represented the main source of
uncertainty. The new calculation also makes use of reweighting in the strange
sea sector, with a corresponding interpolation in the valence sector, to reach
the physical strange quark mass. Renormalization of the four-fermion operator is
performed non-perturbatively using the RI-MOM approach generalized to various
non-exceptional momentum renormalization conditions (for a detailed discussion
of this approach see the plenary talk by Yasumichi Aoki at this
conference~\cite{YAoki}).

The new SBW~09 calculation~\cite{Bae:2009tf}-\cite{Kim:2009ti} (where the
acronym indicates the Seoul, Brookhaven and Washington institutions) uses a
mixed action setup, with HYP-smeared staggered valence fermions, which are
effective at reducing taste-breaking effects, and the asqtad staggered sea
quarks, i.e. the MILC ensembles. Results are obtained at three values of the
lattice spacing, namely the ``coarse'', ``fine'' and ``super-fine'' MILC
ensembles. The chiral extrapolation is performed by using the proper either
SU(3) or SU(2) mixed action staggered ChPT at NLO, with or without adding an
analytical N$^2$LO term. The SU(2) result, which is the one quoted in
table~\ref{tab:BK-ris}, leads eventually to a smaller error. Since the minimum
value of $m_\pi L$ in the simulation is about 2.5, this calculation gets a red
square in table~\ref{tab:BK-ris} for finite volume errors. These errors,
however, have been explicitly investigated by the collaboration and they are
found to be subdominant for $B_K$, smaller than 1\% for the SU(2) analysis. The
main uncertainty in the calculation comes from the evaluation of the
renormalization constant, which is performed in one-loop perturbation theory.
With respect to the older staggered calculation of ref.~\cite{Gamiz:2006sq},
this uncertainty is now significantly reduced, mainly because of the use of the
``fine'' and ``super-fine'' MILC ensembles for which the error coming from the
truncation of the perturbative series, of order $\alpha_s(1/a)^2$, is expected
to be smaller. Nevertheless it still dominates the systematic uncertainty. The
collaboration plans to reduce this uncertainty in various way. One is to work on
a yet finer lattice, another is to use two-loop matching and, finally, by using
the RI-MOM non-perturbative method. This latter approach, which looks the most
promising, has been already successfully applied with staggered fermions to the
renormalization of bilinear quark operators, as discussed by Andrew Lytle at
this conference~\cite{Lytle:2009xm}.

The fourth new result for $B_K$ has been presented at this conference by the ETM
collaboration~\cite{Bertone:2009bu}. This calculation also uses a mixed action
setup, with $N_f=2$ maximally twisted sea quarks and a suitable
Osterwalder-Seiler variant of the twisted mass action for the valence quarks.
This setup simultaneously guarantees the absence of mixing with operators of
wrong chirality, which is usually present with Wilson-like fermions, and
automatic ${\cal O}(a)$-improvement of $B_K$~\cite{Frezzotti:2004wz}. The
calculation is performed at three values of the lattice spacing ($a \simeq
0.07$, 0.09, 0.10 fm) and the physical value of $B_K$ is eventually reached
through a combined SU(2)-chiral and continuum extrapolation. Renormalization of
the four fermion operator is performed non-perturbatively with the RI-MOM
method. Discretization effects in the evaluation of the renormalization
constant, which starts at ${\cal O}(g^2 a^2)$, are further reduced by
subtracting from the RI-MOM four-fermion Green function the leading
contribution, explicitly evaluated in ${\cal O}(a^2)$ one-loop lattice
perturbation theory, as illustrated by Fotos Stylianou in a poster at this
conference~\cite{Constantinou:2010wn}.

\begin{figure}
\begin{center}
\includegraphics[width=.6\textwidth]{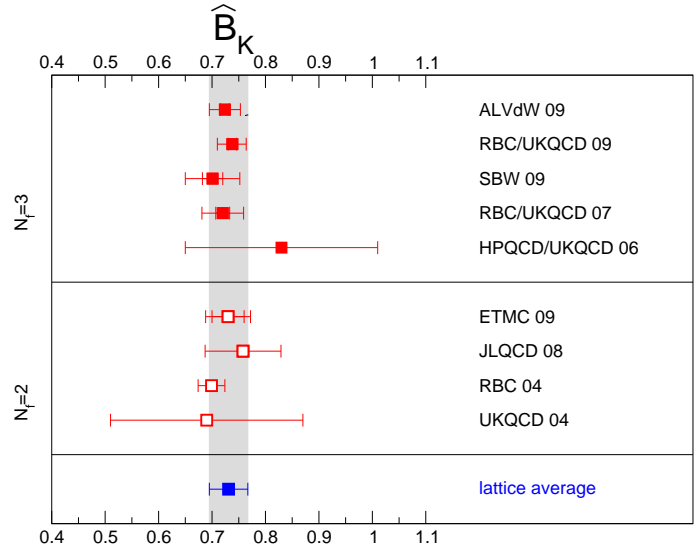}
\end{center}
\caption{{\sl Summary of unquenched lattice results for the RGI parameter $\hat
B_K$ as obtained from simulations with $N_f=2$ and $N_f=3$ dynamical quarks (red
squares). The blue square and the grey band show the average of the lattice
results derived in eq.~(\protect\ref{eq:bkfinal}). This plot has been produced
by FLAG and updated for the lattice conference.}}
\label{fig:bk}
\end{figure}

In order to derive the final lattice average for $B_K$, I consider the three
results which are free of red squares in table~\ref{tab:BK-ris}, namely:
\beq
\renewcommand{\arraystretch}{1.2}
\begin{array}{ll}
\hat B_K = 0.724(8)(28) \quad & [\ N_f=2+1 \ , \ {\rm ALVdW~09}\ ] \\
\hat B_K = 0.738(8)(25) \quad & [\ N_f=2+1 \ , \ {\rm RBC/UKQCD~09}\ ] \\
\hat B_K = 0.730(30)(30) \quad & [\ N_f=2 \ , \ {\rm ETMC~09}\ ] \ .
\end{array}
\renewcommand{\arraystretch}{1.0}
\label{eq:bkresults}
\eeq
The agreement among the above determinations, as well as with the new SBW~09
determination~\cite{Bae:2009tf}-\cite{Kim:2009ti}, is remarkable. It is also
worth noticing that the result obtained with $N_f=2$ dynamical quarks lies in
between the two $N_f=2+1$ determinations, showing that also for $B_K$ the effect
of quenching the strange quark is not visible, within the accuracy currently
reached by lattice calculations. For this reason, I average together the three
results in eq.~(\ref{eq:bkresults}) and quote as the best lattice estimate of
$\hat B_K$
the value
\beq
\setlength{\fboxrule}{2pt}
\fcolorbox{orange}{white}{$\hat B_K = 0.731(7)(35)$} \ .
\label{eq:bkfinal}
\eeq
While the statistical error is evaluated assuming gaussian statistical
uncertainties, the systematic uncertainty quoted in eq.~(\ref{eq:bkfinal}) is
slightly increased with respect to the one quoted by the individual
calculations, to account for the preliminary status of two out of the three
calculations on which the final average of $\hat B_K$ is based. A summary of the
unquenched results for $\hat B_K$ is also presented in fig.~\ref{fig:bk}, which
is an updated version for this conference of a plot produced by
FLAG~\cite{flag}. The average value of $\hat B_K$, given in
eq.~(\ref{eq:bkfinal}), is shown in the plot with a blue square and a grey band.

The $\varepsilon_K$ parameter which controls the amount of indirect CP violation
in $K^0-\bar{K}^0$ mixing, and whose theoretical determination relies on $B_K$,
plays a relevant role in the unitarity triangle analysis, both within and
beyond the Standard Model. Since in the Standard Model the analysis is largely
overconstrained, it can be also used to extract the values of the relevant
hadronic parameters, including $B_K$~\cite{Bona:2006ah}. The latest
determination obtained in this way by the UTfit collaboration, which is
therefore only valid in the Standard Model, is
\beq
(\hat B_K)^{\rm SM}_{\rm UTfit} = 0.87(8)~.
\label{eq:bkutfit}
\eeq
This estimate also takes into account the long distance contributions to both
the absorptive and the dispersive part of the $\Delta S=2$ effective Hamiltonian
evaluated in refs.~\cite{Buras:2008nn,Buras:2010pz} and summarized in the
multiplicative factor $k_\varepsilon=0.94 \pm 0.02$. The Standard Model
prediction of $B_K$ from the unitarity triangle analysis, given in
eq.~(\ref{eq:bkutfit}), shows a tension with the direct lattice
determination~(\ref{eq:bkfinal}), at the level of $1.6\,\sigma$. Whether such a
tension should persist, with the increasing precision of both lattice
calculations and of the unitarity triangle analysis, it could become a clear
signal of physics beyond the Standard Model.

In new physics models, like for instance the supersymmetric extensions of the
Standard Model, the theoretical expression of $\varepsilon_K$ depends in general
on the complete basis of eight $\Delta S=2$ four-fermion
operators~\cite{Ciuchini:1998ix}. Due to parity conservation in the strong
interactions, only five of these operators have independent matrix elements. The
knowledge of these matrix elements is then crucial in order to derive reliable
predictions for $\varepsilon_K$ in the context of specific new physics models.
Lattice calculations for the complete basis of $\Delta S=2$ four-fermion
operators have been only performed so far in the quenched
approximation~\cite{Donini:1999nn,Babich:2006bh,Nakamura:2006eq}, and the
results turn out to be in poor agreement among each other. Very preliminary
results for the full basis of $\Delta S=2$ matrix elements have been presented
at this conference by ETMC~\cite{Bertone:2009bu}. Given their high
phenomenological interest, it would be helpful if such a calculation were also
address by other collaborations using different lattice approaches.

\section{Non leptonic kaon decays}
While not much time (and space) is left to discuss the lattice studies of non
leptonic kaon decays, I cannot conclude this talk without at least mentioning
the extraordinary effort which is being put forward by the RBC/UKQCD
collaboration in addressing this issue. The results obtained last year by the
collaboration have shown that the attractive approach which uses ChPT to relate
the $K\to\pi\pi$ matrix element of interest to the simpler matrix elements of $K
\to \pi$ and $K \to 0$ transitions is affected, in the kaon mass region, by
significantly large chiral corrections. This is an intrinsic uncertainty, which
cannot be avoided. Therefore, the direct calculation of $K\to\pi\pi$ matrix
elements on the lattice must be addressed.

At this conference, three talks have been dedicated to this topic by
RBC/UKQCD~\cite{Liu:2009uw,Lightman:2009cu,Kim:2009fe}. The main indication
is that, while a substantial computational efforts will be required in
order to obtain a reasonably accurate estimate (i.e. at the level of 10-20\%),
the direct calculation of both the $A_0$ and $A_2$ complex amplitudes is however
feasible. The required theoretical and numerical tools include, in particular,
the use of chiral fermions, non-perturbative RI-MOM renormalization and finite
volume methods. 

An exhaustive description of this topic would require a dedicated talk by
itself. For that, I would like to refer the reader to the excellent review given
by Norman Christ at the KAON'09 conference~\cite{Christ:2009ev}.

\section{Summary and outlook}
The number of large unquenched lattice simulations which are being applied to
the study kaon physics is rapidly increasing. This effort has allowed to achieve
in the lattice determination of some (relatively simple) kaon physics
observables an unprecedented accuracy. In this talk, I concentrated most of the
attention in reviewing the lattice results for three of these parameters, namely
the vector semileptonic form factor $f_+(0)$, the ratio of decay constants
$f_K/f_\pi$ and the kaon bag parameter $B_K$. Lattice averages for these
parameters have been given in eqs.~(\ref{eq:fp0final}), (\ref{eq:fkfpifinal})
and (\ref{eq:bkfinal}). Lattice studies of non leptonic kaon decays are
significantly more challenging. Nevertheless, important progress has been
achieved also in this field, and first, reliable results for both the $\Delta
I=1/2$ rule and $\varepsilon'/\varepsilon$ are expected to produced in two ore
three years.

\section*{Acknowledgement}
I would like to thank many colleagues who helped me in preparing this talk, by
providing useful information and/or anticipating to me their results: Claude
Bernard, Norman Christ, Urs Heller, Weonjong Lee, Laurent Lellouch, Chris Kelly,
Matthew Lightman, Bob Mawhinney, Alberto Ramos, Steve Sharpe, Ruth Van de Water,
James Zanotti and all my colleagues in the ETM collaboration. 

A special thank goes to the members of the FLAG working group, Gilberto
Colangelo, Stephan D\"urr, Andreas J\"uttner, Laurent Lellouch, Heiri Leutwyler,
Silvia Necco, Chris Sachrajda, Silvano Simula, Tassos Vladikas, Urs Wenger and
Hartmut Wittig. Much of the material that I have presented in this talk,
including tables and plots, has been prepared in collaboration with them. 

Finally I would like to thank the organizers of Lattice 2009 for inviting me to
give this review and for planning a very enjoyable meeting.



\begin{thebibliography}{99}


\bibitem{flag}
  G.~Colangelo {\it et al.}  [FLAG Working Group of FLAVIANET],
  ``Review of lattice results concerning low energy particle physics'',
  in preparation. See also G.~Colangelo, talk given at EuroFlavour 2009,
  {\tt http://www.ba.infn.it/indico/conferenceDisplay.py?confId=49}

\bibitem{Baron:2009zq}
  R.~Baron {\it et al.},
  arXiv:0911.5244 [hep-lat].

\bibitem{flavianet}
  M.~Antonelli {\it et al.}  [FlaviaNet Working Group on Kaon Decays],
  arXiv:0801.1817 [hep-ph].

\bibitem{Hardy:2008gy}
  J.~C.~Hardy and I.~S.~Towner,
  Phys.\ Rev.\  C {\bf 79} (2009) 055502
  [arXiv:0812.1202 [nucl-ex]].

\bibitem{ag}
  M.~Ademollo and R.~Gatto,
  Phys.\ Rev.\ Lett.\  {\bf 13} (1964) 264.

\bibitem{gl}
  J.~Gasser and H.~Leutwyler,
  Nucl.\ Phys.\  B {\bf 250} (1985) 517.

\bibitem{lr}
  H.~Leutwyler and M.~Roos,
  Z.\ Phys.\  C {\bf 25} (1984) 91.


\bibitem{Becirevic:2004ya} 
  D.~Becirevic {\it et al.},
  Nucl.\ Phys.\  B {\bf 705} (2005) 339
  [arXiv:hep-ph/0403217].

\bibitem{Boyle:2007qe} 
  P.~A.~Boyle {\it et al.},
  Phys.\ Rev.\ Lett.\  {\bf 100} (2008) 141601
  [arXiv:0710.5136 [hep-lat]].

\bibitem{Lubicz:2009ht}
  V.~Lubicz, F.~Mescia, S.~Simula and C.~Tarantino [ETM Collaboration],
  Phys.\ Rev.\  D {\bf 80} (2009) 111502
  [arXiv:0906.4728 [hep-lat]].

\bibitem{Brommel:2007wn} 
  D.~Brommel {\it et al.}  [QCDSF collaboration],
  PoS {\bf LAT2007} (2007) 364
  [arXiv:0710.2100 [hep-lat]].

\bibitem{Dawson:2006qc} 
  C.~Dawson, T.~Izubuchi, T.~Kaneko, S.~Sasaki and A.~Soni,
  Phys.\ Rev.\  D {\bf 74} (2006) 114502
  [arXiv:hep-ph/0607162].

\bibitem{Tsutsui:2005cj} 
  N.~Tsutsui {\it et al.}  [JLQCD Collaboration],
  PoS {\bf LAT2005} (2006) 357
  [arXiv:hep-lat/0510068].


\bibitem{jzanotti}
  RBC and UKQCD Collaborations, talk given by J.~Zanotti at this conference.

\bibitem{DiVita:2009by}
  S.~Di Vita, B.~Haas, F.~Mescia, V.~Lubicz, S.~Simula and C.~Tarantino
  [ETM Collaboration],
  PoS {\bf LAT2009} (2009) 257
  [arXiv:0910.4845 [hep-lat]].

\bibitem{Boyle:2007wg}
  P.~A.~Boyle, J.~M.~Flynn, A.~Juttner, C.~T.~Sachrajda and J.~M.~Zanotti,
  JHEP {\bf 0705} (2007) 016
  [arXiv:hep-lat/0703005].

\bibitem{Guadagnoli:2005be}
  D.~Guadagnoli, F.~Mescia and S.~Simula,
  Phys.\ Rev.\  D {\bf 73} (2006) 114504
  [arXiv:hep-lat/0512020].

\bibitem{Sachrajda:2004mi}
  C.~T.~Sachrajda and G.~Villadoro,
  Phys.\ Lett.\  B {\bf 609} (2005) 73
  [arXiv:hep-lat/0411033].

\bibitem{Roessl:1999iu}
  A.~Roessl,
  Nucl.\ Phys.\  B {\bf 555} (1999) 507
  [arXiv:hep-ph/9904230].

\bibitem{Allton:2008pn}
  C.~Allton {\it et al.}  [RBC-UKQCD Collaboration],
  Phys.\ Rev.\  D {\bf 78} (2008) 114509
  [arXiv:0804.0473 [hep-lat]].

\bibitem{Flynn:2008tg}
  J.~M.~Flynn and C.~T.~Sachrajda  [RBC and UKQCD Collaborations],
  Nucl.\ Phys.\  B {\bf 812} (2009) 64
  [arXiv:0809.1229 [hep-ph]].

\bibitem{Scholz:2009yz}
  E.~E.~Scholz,
  PoS {\bf LAT2009} (2009) 005
  arXiv:0911.2191 [hep-lat].

\bibitem{Becirevic:2005py}
  D.~Becirevic, G.~Martinelli and G.~Villadoro,
  Phys.\ Lett.\  B {\bf 633} (2006) 84
  [arXiv:hep-lat/0508013].

\bibitem{Boyle:2010bh}
  P.~A.~Boyle {\it et al.},
  arXiv:1004.0886 [Unknown].

\bibitem{simula} S.~Simula, private communication.

\bibitem{Kastner:2008ch}
  A.~Kastner and H.~Neufeld,
  Eur.\ Phys.\ J.\  C {\bf 57} (2008) 541
  [arXiv:0805.2222 [hep-ph]].

\bibitem{Cirigliano:2005xn}
  V.~Cirigliano, G.~Ecker, M.~Eidemuller, R.~Kaiser, A.~Pich and J.~Portoles,
  JHEP {\bf 0504} (2005) 006
  [arXiv:hep-ph/0503108].

\bibitem{Jamin:2004re}
  M.~Jamin, J.~A.~Oller and A.~Pich,
  JHEP {\bf 0402} (2004) 047
  [arXiv:hep-ph/0401080].

\bibitem{Bijnens:2003uy}
  J.~Bijnens and P.~Talavera,
  Nucl.\ Phys.\  B {\bf 669} (2003) 341
  [arXiv:hep-ph/0303103].


\bibitem{ALVdW09} 
 C.~Aubin, J.~Laiho and R.~S.~Van de Water, talk given by J.~Laiho at 
 Chiral Dynamics 2009, Bern, Switzerland, July 6-10, 2009.
 http://www.chiral09.unibe.ch/

\bibitem{Ramos:2010ar}
  A.~Ramos and f.~M.~collaboration,
  PoS {\bf LAT2009} (2009) 259
  [arXiv:1002.1665 [hep-lat]].

\bibitem{Durr:2010hr}
  S.~Durr {\it et al.},
  arXiv:1001.4692 [hep-lat].

\bibitem{Mawhinney:2009jy} 
  R.~Mawhinney  [RBC Collaboration and UKQCD Collaboration],
  PoS {\bf LAT2009} (2009) 081
  [arXiv:0910.3194 [hep-lat]].
CI,LAT2009,081;

\bibitem{Bazavov:2009tw} 
  A.~Bazavov {\it et al.}  [The MILC Collaboration],
  PoS {\bf LAT2009} (2009) 079
  [arXiv:0910.3618 [hep-lat]].

\bibitem{Bazavov:2009bb} 
  A.~Bazavov {\it et al.},
  arXiv:0903.3598 [hep-lat].

\bibitem{JLQCD:2009sk} 
  J.~Noaki {\it et al.}  [JLQCD and TWQCD Collaborations],
  PoS {\bf LAT2009} (2009) 096
  [arXiv:0910.5532 [hep-lat]].

\bibitem{Aoki:2008sm}
  S.~Aoki {\it et al.}  [PACS-CS Collaboration],
  Phys.\ Rev.\  D {\bf 79} (2009) 034503
  [arXiv:0807.1661 [hep-lat]].

\bibitem{Follana:2007uv} 
  E.~Follana, C.~T.~H.~Davies, G.~P.~Lepage and J.~Shigemitsu  [HPQCD
  Collaboration and UKQCD Collaboration],
  Phys.\ Rev.\ Lett.\  {\bf 100} (2008) 062002
  [arXiv:0706.1726 [hep-lat]].

\bibitem{Beane:2006kx} 
  S.~R.~Beane, P.~F.~Bedaque, K.~Orginos and M.~J.~Savage,
  Phys.\ Rev.\  D {\bf 75} (2007) 094501
  [arXiv:hep-lat/0606023].

\bibitem{Aubin:2004fs} 
  C.~Aubin {\it et al.}  [MILC Collaboration],
  Phys.\ Rev.\  D {\bf 70} (2004) 114501
  [arXiv:hep-lat/0407028].

\bibitem{Blossier:2009bx} 
  B.~Blossier {\it et al.},
  JHEP {\bf 0907} (2009) 043
  [arXiv:0904.0954 [hep-lat]].

\bibitem{Blossier:2007vv} 
  B.~Blossier {\it et al.}  [ETM Collaboration],
  JHEP {\bf 0804} (2008) 020
  [arXiv:0709.4574 [hep-lat]].

\bibitem{QCDSF07} 
  G.~Schierholz, talk given at Lattice 2007, Regensburg, Germany,
  http://www.physik.uni-regensburg.de/lat07/hevea/schierholz.pdf


\bibitem{Rosner:2010ak}
  J.~L.~Rosner and S.~Stone,
  arXiv:1002.1655 [hep-ex].


\bibitem{Sharpe:1996ih}
  S.~R.~Sharpe,
  Nucl.\ Phys.\ Proc.\ Suppl.\  {\bf 53} (1997) 181
  [arXiv:hep-lat/9609029].


\bibitem{Lellouch:2000bm}
  L.~Lellouch,
  Nucl.\ Phys.\ Proc.\ Suppl.\  {\bf 94} (2001) 142
  [arXiv:hep-lat/0011088].

\bibitem{Dawson:2005za}
  C.~Dawson,
  PoS {\bf LAT2005}, 007 (2006).

\bibitem{Lellouch:2009fg}
  L.~Lellouch,
  PoS {\bf LATTICE2008} (2009) 015
  [arXiv:0902.4545 [hep-lat]].


\bibitem{Aubin:2009jh} 
  C.~Aubin, J.~Laiho and R.~S.~Van de Water,
  Phys.\ Rev.\  D {\bf 81} (2010) 014507
  [arXiv:0905.3947 [hep-lat]].

\bibitem{Kelly:2009fp} 
  C.~Kelly, P.~A.~Boyle and C.~T.~Sachrajda  [RBC Collaboration and UKQCD
                  Collaboration],
  PoS {\bf LAT2009} (2009) 087
  [arXiv:0911.1309 [hep-lat]].

\bibitem{Bae:2009tf} 
  T.~Bae {\it et al.},
  PoS {\bf LAT2009} (2009) 261
  [arXiv:0910.5576 [hep-lat]].

\bibitem{Kim:2009te}
  H.~J.~Kim {\it et al.},
  PoS {\bf LAT2009} (2009) 262
  [arXiv:0910.5573 [hep-lat]].

\bibitem{Yoon:2009th}
  B.~Yoon {\it et al.},
  PoS {\bf LAT2009} (2009) 263
  [arXiv:0910.5581 [hep-lat]].

\bibitem{Kim:2009ti}
  J.~Kim {\it et al.},
  PoS {\bf LAT2009} (2009) 264
  [arXiv:0910.5583 [hep-lat]].

\bibitem{Antonio:2007pb} 
  D.~J.~Antonio {\it et al.}  [RBC Collaboration and UKQCD Collaboration],
  Phys.\ Rev.\ Lett.\  {\bf 100} (2008) 032001
  [arXiv:hep-ph/0702042].

\bibitem{Gamiz:2006sq} 
  E.~Gamiz, S.~Collins, C.~T.~H.~Davies, G.~P.~Lepage, J.~Shigemitsu and
  M.~Wingate      [HPQCD Collaboration and UKQCD Collaboration],
  Phys.\ Rev.\  D {\bf 73} (2006) 114502
  [arXiv:hep-lat/0603023].

\bibitem{Bertone:2009bu}
  V.~Bertone {\it et al.}  [ETM Collaboration],
  PoS {\bf LAT2009} (2009) 258
  [arXiv:0910.4838 [hep-lat]].

\bibitem{Aoki:2008ss} 
  S.~Aoki {\it et al.}  [JLQCD Collaboration],
  Phys.\ Rev.\  D {\bf 77} (2008) 094503
  [arXiv:0801.4186 [hep-lat]].

\bibitem{Aoki:2004ht} 
  Y.~Aoki {\it et al.},
  Phys.\ Rev.\  D {\bf 72} (2005) 114505
  [arXiv:hep-lat/0411006].

\bibitem{Flynn:2004au} 
  J.~M.~Flynn, F.~Mescia and A.~S.~B.~Tariq  [UKQCD Collaboration],
  JHEP {\bf 0411} (2004) 049
  [arXiv:hep-lat/0406013].


\bibitem{Bar:2005tu}
  O.~Bar, C.~Bernard, G.~Rupak and N.~Shoresh,
  Phys.\ Rev.\  D {\bf 72} (2005) 054502
  [arXiv:hep-lat/0503009].

\bibitem{Aubin:2006hg}
  C.~Aubin, J.~Laiho and R.~S.~Van de Water,
  Phys.\ Rev.\  D {\bf 75} (2007) 034502
  [Erratum-ibid.\  D {\bf 79} (2009) 079904]
  [arXiv:hep-lat/0609009].

\bibitem{YAoki}
  Y.~Aoki, plenary talk at this conference,
{\tt http://rchep.pku.edu.cn/workshop/lattice09/Jul 2028 20TUE/Aoki.pdf}

\bibitem{Lytle:2009xm}
  A.~T.~Lytle,
  PoS {\bf LAT2009} (2009) 202
  [arXiv:0910.3721 [hep-lat]].

\bibitem{Frezzotti:2004wz}
  R.~Frezzotti and G.~C.~Rossi,
  JHEP {\bf 0410} (2004) 070
  [arXiv:hep-lat/0407002].

\bibitem{Constantinou:2010wn}
  M.~Constantinou, V.~Lubicz, H.~Panagopoulos, A.~Skouroupathis and
F.~Stylianou,
  arXiv:1001.1241 [hep-lat].

\bibitem{Bona:2006ah}
  M.~Bona {\it et al.}  [UTfit Collaboration],
  JHEP {\bf 0610} (2006) 081
  [arXiv:hep-ph/0606167].

\bibitem{Buras:2008nn}
  A.~J.~Buras and D.~Guadagnoli,
  Phys.\ Rev.\  D {\bf 78} (2008) 033005
  [arXiv:0805.3887 [hep-ph]].

\bibitem{Buras:2010pz}
  A.~J.~Buras, D.~Guadagnoli and G.~Isidori,
  arXiv:1002.3612 [Unknown].

\bibitem{Ciuchini:1998ix}
  M.~Ciuchini {\it et al.},
  JHEP {\bf 9810} (1998) 008
  [arXiv:hep-ph/9808328].

\bibitem{Donini:1999nn}
  A.~Donini, V.~Gimenez, L.~Giusti and G.~Martinelli,
  Phys.\ Lett.\  B {\bf 470} (1999) 233
  [arXiv:hep-lat/9910017].

\bibitem{Babich:2006bh}
  R.~Babich, N.~Garron, C.~Hoelbling, J.~Howard, L.~Lellouch and C.~Rebbi,
  Phys.\ Rev.\  D {\bf 74} (2006) 073009
  [arXiv:hep-lat/0605016].

\bibitem{Nakamura:2006eq}
  Y.~Nakamura {\it et al.}  [CP-PACS Collaboration],
  PoS {\bf LAT2006} (2006) 089
  [arXiv:hep-lat/0610075].


\bibitem{Liu:2009uw}
  Q.~Liu  [RBC Collaboration and UKQCD Collaboration],
  arXiv:0910.2658 [hep-lat].

\bibitem{Lightman:2009cu}
  M.~Lightman and E.~Goode  [RBC Collaboration and UKQCD Collaboration],
  PoS {\bf LAT2009} (2009) 254
  [arXiv:0912.1667 [Unknown]].

\bibitem{Kim:2009fe}
  C.~Kim and N.~H.~Christ,
  PoS {\bf LAT2009} (2009) 255
  [arXiv:0912.2936 [hep-lat]].

\bibitem{Christ:2009ev}
  N.~H.~Christ  [RBC Collaboration and UKQCD Collaboration],
  arXiv:0912.2917 [hep-lat].

\end{thebibliography}
\end{document}